# Structured Spreadsheet Modelling and Implementation with Multiple Dimensions - Part 1: Modelling

Paul Mireault
Founder, SSMI International
Honorary Professor, HEC Montréal
Paul.Mireault@SSMI.International

**ABSTRACT**

*Dimensions are an integral part of many models we use every day. Without thinking about it, we frequently use the time dimension: many financial and accounting spreadsheets have columns representing months or years. Representing a second dimension is often done by repeating blocs of formulas in a worksheet of creating multiple worksheets with the same structure.*

## 1 Introduction

Most organizations deal with dimensions, without calling them as such. For example:

- Products, product categories, product types.
- Clients, client types, client status.
- Markets or sectors, like education or health.
- Locations. They may be geographical (such as countries, continents, regions) or specific (such as manufacturing plants, warehouses) or arbitrary (such as sales regions).

In this paper, we will first present some examples of multidimensional spreadsheets. Then, we will do a brief summary of the conceptual modelling methodology we use to represent the problem we wish to solve with a spreadsheet. We will then present basic concepts of dimensions, variables and multidimensional expressions. We conclude with a case study and describe its complete multidimensional model.

## 2 Examples of multidimensional spreadsheets

Microsoft Excel has a tool called Pivot Table that can help the spreadsheet developer present a multidimensional dataset in a two-dimensional table, using rows and columns to represent more than one dimension. While Pivot Tables are good for presenting data, they are less suited for presenting business spreadsheets. The principal reason is that Pivot Tables require that their source is organized vertically as tables: each column represents a variable, and the rows represent the repeated values. The spreadsheets we are interested in are organized horizontally: the rows represent variables and the columns are the repeated values. We could transpose an horizontal structure into a vertical one, but this extra step does not alleviate Pivot Table's other shortcoming. The major reason we feel that Pivot Tables are not suitable for spreadsheet that represent a model used to analyse scenarios, as opposed to a spreadsheet containing data, is that they do not update their results when their base data changes. For that reason, the results produced by Pivot Tables cannot be used in the calculations of other variables.

Even though a spreadsheet has two dimensions, rows and columns, it usually represents only one dimension. Most business spreadsheet use rows for variables, leaving the columns for one dimension, like the Time dimension.

One approach is to create one worksheet for each instance of a dimension and implementing the other dimensions inside those worksheets. For example, if the dimension is Region, there could be four worksheets for North, East, West and South. One could then build a fifth worksheet with consolidating formulas.

This method is proposed by (Sartain, 2014) where the author describes using 13 worksheets, one for each month and one for consolidation, and each worksheet assigns expense variables, in rows, to different persons, in columns (see Figure 1). The author also describes the maintenance task of adding or removing an account, which involves performing the same operation 13 times, once for each worksheet. She also strongly suggests, when adding an account, to add the row somewhere in the middle of the other accounts to make sure that the subtotal row includes it in its calculation.

*Figure 1 Multidimensional spreadsheet example from (Sartain, 2014)*

(Brandewinder, 2008) has a spreadsheet with three dimensions: *Quarter*, *Product* and *Region* (see Figure 2). The *Product* dimension is presented as different worksheets, the *Quarter* dimension as columns and the *Region* dimension as blocs of repeated formulas.

*Figure 2 Multidimensional spreadsheet example from (Brandewinder, 2008)*

One can use an entire worksheet to represent one two-dimensional variable. Figure 3 shows an unpublished example where the two-dimensional variable **Border Right Indic** is implemented in its own worksheet.

*Figure 3 Example of a worksheet used for one two-dimensional variable*

(Savage, 1997) describes two important problems with using dimensions in spreadsheets. First is *scalability*, which involves changing the cardinality of a dimension. He concludes that spreadsheets rarely scale well. Second is *hyper-scalability*, which involves changing the dimensions themselves, such as adding more dimensions. His conclusion is, succinctly, "Forget it".

Multi-dimensional spreadsheets have also been used in specific optimization problems. (Kumar, 2014) describes a course scheduling problem with three dimensions: faculty, course and timeslot. A textbook by (Powell & Baker, 2013) presents many classic Management Science problems such as the Network Flow, the Assignment and the Traveling Salesman. While they present some multi-dimensional problems, their spreadsheets are specific to each problem.

## 3 The Conceptual Model

In Information Systems development, the stage were the requirements are specified produces the *conceptual model*. The conceptual model describes what the system must do, with little reference to the technology that will be used for the implementation.

(Grossman & Özlük, 2010) in their study of three spreadsheet engineering methodologies found that two of them do not discuss modeling and the other requires a detailed output specification.

Other researchers described building a conceptual model before implementing the spreadsheet, even though they did not call it *conceptual modelling*. The Jackson Structured Diagram, a diagraming technique based on programming concepts, has been proposed by (Knight, Chadwick, & Rajalingham, 2000). Their diagram has some similarities with the simple Formula Diagram of (Mireault, 2017), but they do not show how to extend it to a one dimension model. (Powell & Baker, 2013) use Influence Charts to model a problem and give general advice on how to implement it in a spreadsheet. While their examples show a one-dimension spreadsheet, with Quarters, their Influence Chart does not show which variables belong to the Quarter dimension.

### 3.1 The Formula Diagram of the SSMI Methodology

(Mireault, 2017) presents a methodology for developing spreadsheets based, primarily, on following the process used in information systems development, where the requirement specifications is separate from the implementation. The Structured Spreadsheet Modelling and Implementation (SSMI) methodology consist of building a conceptual model of the spreadsheet's variables and their formulas before doing the implementation. The conceptual model is composed of a Formula Diagram (Figure 4) and a Formula List (Figure 5), and they are used later to do the implementation of the spreadsheet.

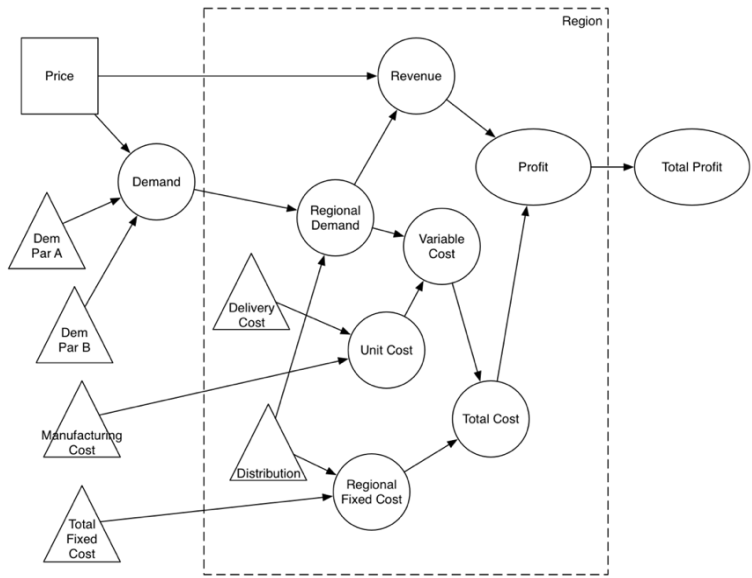

*Figure 4 Example of a Formula Diagram, taken from (Mireault, 2017)*

The Formula Diagram uses the following symbols:

- Triangles and squares represent data values. The squares are used for input values, data that the developer wants to implement in an Interface worksheet to allow the user to easily modify its value. The other data, the triangles, represent data that don't change often and will be implemented in their own specific data worksheets.
- Circles and ovals represent calculated variables. The ovals are used for results that the developer wants to display in the Interface worksheet, close to the input data so the user can quickly see the impact of changing an input value. The circles are variables of less interest to the user and are implemented in their own specific model worksheets.
- Arrows indicate which variables are involved in the calculation of the variable receiving them.
- The dash-bordered box represents a dimension, also called entity. All the variables appearing within the box have multiple values, one value for each instance of the repeated entity. For example, if we have three regions, then the variable **Regional Demand** has three values. All the variables appearing outside the dashed box have a single value.

| Variable | Type | Definition |
|---|---|---|
| *Price* | Input | *(To be set by user)* |
| *Profit* | Output, Region | *Revenue – Total Cost* |
| *DemParA* | Data | *376,000* |
| *DemParB* | Data | *1.009* |
| *Fixed Cost* | Data | *$2,500,000* |
| *Manufacturing Cost* | Data | *$120* |
| *Distribution* | Data, Region | *48%, 23%, 29%* |
| *Delivery Cost* | Data, Region | *$50, $80, $60* |
| *Total Demand* | Calculated | *DemParA * DemParB^–Price* |
| *Regional Demand* | Calculated, Region | *Total Demand * Distribution* |
| *Total Cost* | Calculated, Region | *Regional Fixed Cost + Variable Cost* |
| *Regional Fixed Cost* | Calculated, Region | *Fixed Cost * Distribution* |
| *Variable Cost* | Calculated, Region | *Regional Demand * Unit Cost* |
| *Unit Cost* | Calculated, Region | *Manufacturing Cost + Delivery Cost* |
| *Revenue* | Calculated, Region | *Regional Demand * Price* |
| *Total Profit* | Output | *SUM(Profit)* |

*Figure 5 Example of a Formula List*

While the Formula Diagram gives a global view of the model, the corresponding Formula List gives a detailed view, with all the formulas written in an Excel-like form, using variable names.

The Formula Diagram is inspired from the Influence Diagram, as presented in (Bodily, 1985). The Influence Diagram has a richer set of modeling concepts, such as uncertainty in the values of data variables and uncertainty in the formulas of calculated variables. But the Influence Diagram has no representation of groups of repeating variables, which the Formula Diagram represents with a dash-bordered box.

## 4 Multidimensional modelling concepts

At this point, we invite the reader to read the case study presented in the appendix so that they can get a better appreciation of the concepts we present in this section.

### 4.1 Dimensions

A dimension is a set of values that serve to characterize a specific value. The set of values form a *partition*. A partition, in set theory, represent subsets whose intersections, taken two by two, are null, and whose union is the universal set, that is the set of all values. In plain language, it means that there is no overlap and all possibilities are covered.

For example, if we use the dimension *Region* to characterize clients and we have the set of values {Mountain, Valley, Lake}, a client must belong to one of the regions (all possibilities are covered) and cannot belong to two regions (no overlap).

### 4.2 Dimension sets

A *dimension set* is a set comprised of 0 or more dimension, and a variable belongs to a specific dimension set. Often, the variable name we use gives a clue to the dimension set it belongs to: the variable named **Monthly Production** belongs to the dimension set (*Month*) and the variable **Monthly Regional Sales** belongs to the dimension set (*Month*, *Region*).

We will say that a variable belonging to the empty, (), dimension set is *dimensionless*. We will also say that dimension sets composed of only one dimension are *basic*. Finally, the dimension set composed of all the dimensions is called the *full* dimension set.

If we have $n$ dimensions, then we have $2^n$ possible dimension sets, ranging from the empty set to the set of all dimensions. Thus, when we have only one dimension, like *Time*, a variable either belongs to the (*Time*) dimension set or is dimensionless. If we have two dimensions, like *Month* and *Region*, a variable either belongs to the (*Month*, *Region*) dimension set, the (*Month*) dimension set, the (*Region*) dimension set or the () dimension set.

### 4.3 Defining variables

In usual mathematical notation, a variable's dimension set is indicated by subscripts. Thus, the two variables described above would be written like this: **Monthly Production**$_{Month}$ and **Monthly Regional Sales**$_{Month,Region}$. It is redundant to specify the dimension set in the variable's name and in the subscripts: we will only do so in this section because we want to make sure that the dimensions are clear.

There are mathematical rules to remember when dealing with expressions involving variables of different dimension sets. We usually apply them without thinking about it because they are common sense. We will describe the rules and show how they are represented in a Formula Diagram and a Formula List.

Rule 1: The dimension set of a formula is the union of the dimension sets of all the variables that are part of its definition.

Example 1:

- **Unit Production Cost** is of dimension set (*Product*).
- **Unit Delivery Cost** is of dimension set (*Region*).

- **Unit Cost = Unit Production Cost + Unit Delivery Cost** is thus of dimension set (*Product, Region*).
- The mathematical representation of the formula is:

Unit Cost $_{Product, Region}$ = Unit Production Cost $_{Product}$ + Unit Delivery Cost $_{Region}$

- Figure 6 illustrates how this variable definition is shown in a Formula Diagram.

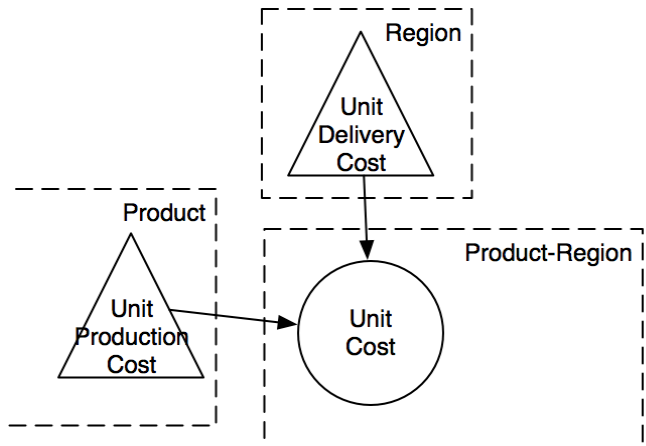

*Figure 6 Defining a two-dimensional variable from two one-dimensional variables*

Example 2:

- **Annual Sector-Product Unit Sales** is of dimension set (*Sector, Product*).
- **Monthly Sales Distribution per Sector** is of dimension set (*Month, Sector*).
- **Monthly-Sector-Product Unit Sales = Annual Sector-Product Unit Sales * Monthly Sales Distribution per Sector** is thus of dimension set (*Month, Sector, Product*).
- The mathematical representation of the formula is:

Monthly-Sector-Product Unit Sales $_{Month, Sector, Product}$
    = Annual Sector-Product Unit Sales $_{Sector, Product}$
    + Monthly Sales Distribution per Sector $_{Month, Sector}$

- This is shown in Figure 7.

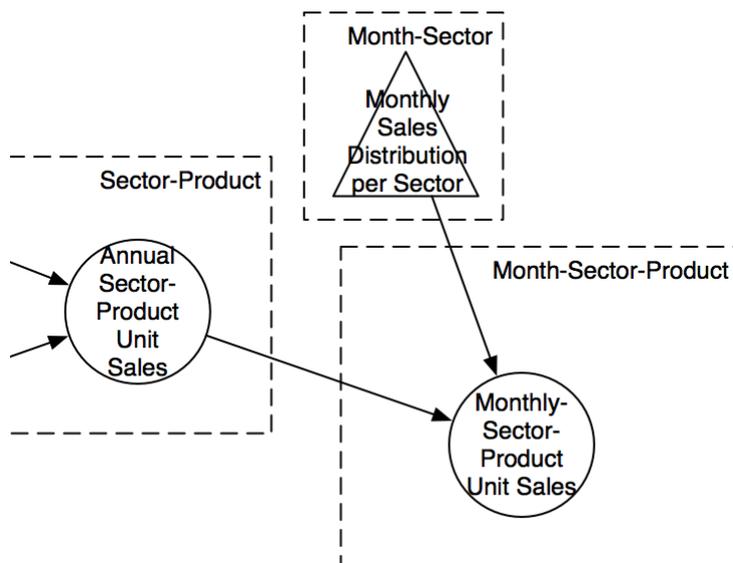

*Figure 7 Defining a three-dimensional variable from two two-dimensional variables*

Rule 2: Besides aggregation, a variable can only be defined with variables having a dimension set that is a subset of its own.

Example:

- **Monthly-Sector-Product Unit Sales** is of dimension set (*Month*, *Sector*, *Product*)
- It can be defined with variables of dimension sets (*Month*, *Sector*, *Product*), (*Month*, *Sector*), (*Month*, *Product*), (*Sector*, *Product*), (*Month*), (*Sector*), (*Product*) and ().
- It cannot be defined with variables of dimension sets (*Month*, *Region*) or (*Product*, *Region*)
- Figure 6 and Figure 7 are also illustrations of this rule.

Rule 3: In the case of an aggregation, a variable can only be defined with a variable having a dimension set that is a superset of its own.

Example 1:

- **Regional Unit Sales** is of dimension (*Region*).
- It can be aggregated from a variable of dimension sets (*Month*, *Product*, *Region*), (*Sector*, *Region*) or (*Month*, *Region*).
- It cannot be aggregated from a variable of dimension sets (*Month*, *Sector*) or (*Product*).
- In the Formula List, we would write it as =**SUM(Monthly-Product-Region Unit Sales)** or =**SUM(Monthly-Region Unit Sales).** Mathematically, both formulas are equivalent.
- The mathematical representations of the two formulas are:

$$\text{Regional Unit Sales}_{Region} = \sum_{\substack{Month \\ Product}} \text{Monthly-Product-Region Unit Sales}_{Month,\ Product,\ Region}$$

or

$$\text{Regional Unit Sales}_{Region} = \sum_{Month} \text{Monthly-Region Unit Sales}_{Month,\ Region}$$

- Figure 8 shows how we would present the two formulas in the Formula Diagram.

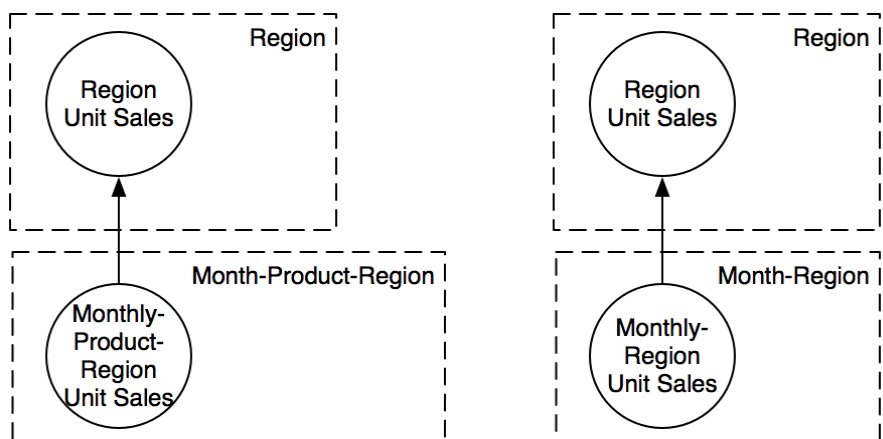

*Figure 8 Two ways to define the same aggregate variable*

Example 2:

- **Regional-Product Unit Sales** is of dimension set (*Product*, *Region*).
- It can be aggregated from a variable of dimension sets (*Month*, *Product*, *Region*), (*Sector*, *Product*, *Region*) or (*Month*, *Sector*, *Product*, *Region*).
- The mathematical representation of the formula is:

$$\text{Regional-Product Unit Sales}_{Product,\ Region} = \sum_{Month} \text{Monthly-Product-Region Unit Sales}_{Month,\ Product,\ Region}$$

- It cannot be aggregated from a variable of dimension sets (*Month*, *Region*) or (*Product*).

Example 3:

- **Total Unit Sales** is of dimension set ().
- It can be aggregated from a variable of any dimension set.
- In the Formula List, we would write it as =**SUM(Monthly-Sector-Product-Region Unit Sales)** or =**SUM(Monthly-Product-Region Unit Sales).** Mathematically, both formulas are equivalent.
- The mathematical representations of the two formulas are:

$$\text{Total Unit Sales} = \sum_{\substack{Month \\ Sector \\ Product \\ Region}} \text{Monthly-Sector-Product-Region Unit Sales}_{\text{Month, Sector, Product, Region}}$$

or

$$\text{Total Unit Sales} = \sum_{\substack{Month \\ Product \\ Region}} \text{Monthly-Product-Region Unit Sales}_{\text{Month, Product, Region}}$$

In the Formula Diagram, the dimensionless variable is shown outside of any box, as illustrated in Figure 9.

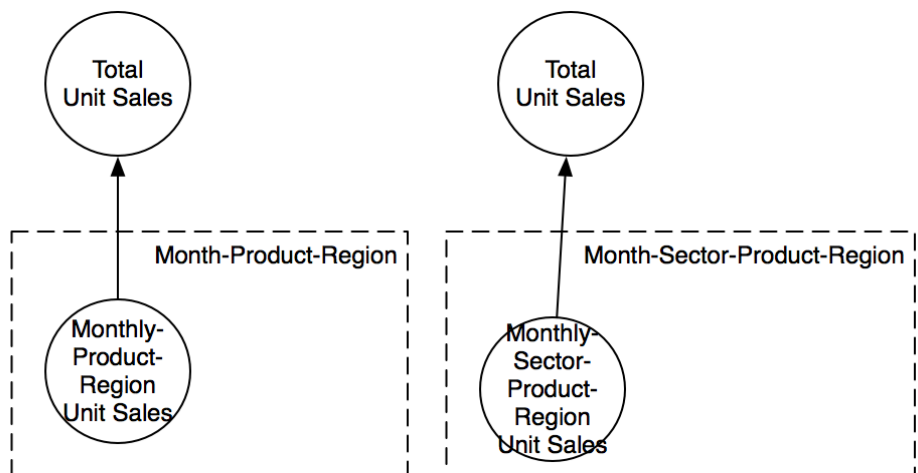

*Figure 9 Defining a dimensionless variable as an aggregate of a multidimensional variable*

## 5    Conclusion

In this paper, we extended the one-dimension conceptual model of (Mireault, 2017) to model a multidimensional problem. By building the conceptual model before doing the implementation, the spreadsheet developer is not trying to solve two different problems at the same time: *How do I calculate this variable?* and *How do I implement this in my spreadsheet?*

An experiment by (O'Donnel, 2001) showed that using a diagramming technique, an Influence Diagram in this case, did not take significantly more time to produce the final spreadsheet and those spreadsheets had significantly less errors of the type "omitted factors" than those of the control group. The problem submitted to the test subjects was a relatively simple one, with one time dimension of two periods. It would be interesting to reproduce the experiment with a more complex multi-dimensional problem such as the one presented in the Appendix.

Part 2 of this paper will present a structured methodology to implement the multidimensional model.

**Appendix – Case Study**

In this section, we present a pedagogical case study to illustrate the concepts presented in this paper. The solution is not unique: there are many ways of calculating some variables, as illustrated above.

**The Acme TechnoWidget Company**

The Acme TechnoWidget Company produces and sells widgets. It produces two products (a Standard widget and a Deluxe widget) and its salesforce is assigned to four major sectors: Government, military, education and private.

Market research has established that the annual demand for widgets depends on each sector's Standard widget price. The Pricing Director explains:

*We start by setting a global base price. Then, for each sector, we tell our salesforce that they can offer a rebate. For instance, we offer a 70% rebate to the education sector and it's 10% for the private sector because purchases are usually made by researchers with limited funds. The military sector gets a 20% rebate and the government 40%. This is not made public: all our price lists show the base price, but our clients in each sector are aware of the rebate they can get.*

*Each sector reacts differently to a change of price. We consulted with a market research expert and she came up with multiple demand functions, one for each sector. The demand function estimates a sector's annual demand for a given base price. The demand function has the form $B/\text{Price}^A$. The parameters A and B are different for each sector, and* Price *is the sector's price, after the rebate. This table shows the values the expert gave us:*

| Sector | Government | Military | Private | Education |
|---|---|---|---|---|
| *Rebate Percentage* | *40%* | *20%* | *10%* | *70%* |
| *DemParA* | *3.593437587* | *3.46315031* | *3.187228762* | *4.114496316* |
| *DemParB* | *22858963442* | *22858963442* | *22858963442* | *22858963442* |

*The price of the Deluxe widget is 45% higher than the Standard widget.*

The Sales Manager explains the sales pattern:

*The annual demand of each Sector is split between the Standard and Deluxe product types, but the distribution is very different in each sector. For instance, in the education sector, with its limited funds, the split is 80%-20% and it is 25%-75% in the military sector. I guess these guys always go for the best, and they have higher budgets. The distribution is 65%-35% for the government sector and 40%-60% for the private sector. The ratios are then applied to the sector's annual demand to get the annual demand by product.*

*Another interesting pattern is the distribution of sales during the year. We noticed that our clients buy more just before the end of their fiscal year, when some want to spend their budget surpluses, and the beginning, when others have new funds allotted. Each sector has a different pattern, and we noticed that it is pretty stable year after year.*

|  | Government | Military | Private | Education |
|---|---|---|---|---|
| Jan | 9% | 8% | 12% | 6% |
| Feb | 10% | 9% | 11% | 8% |
| Mar | 12% | 10% | 9% | 9% |
| Apr | 12% | 12% | 7% | 10% |
| May | 11% | 13% | 6% | 12% |
| Jun | 9% | 11% | 4% | 12% |
| Jul | 7% | 9% | 5% | 11% |
| Aug | 6% | 7% | 6% | 9% |
| Sep | 5% | 6% | 8% | 7% |
| Oct | 5% | 4% | 9% | 6% |
| Nov | 6% | 5% | 11% | 5% |
| Dec | 8% | 6% | 12% | 5% |
| Total | 100% | 100% | 100% | 100% |

*Sales to a sector are not uniformly distributed by region. For example, there are more universities in the South-West than in the West. The following table shows the distribution of a sector's sales by region. With it, we can calculate the expected monthly sales per product per region, which helps our Logistics Department do its planning.*

| | Government | Military | Private | Education |
|---|---|---|---|---|
| N | 25% | 52% | 22% | 24% |
| SE | 18% | 13% | 21% | 15% |
| SW | 18% | 18% | 17% | 32% |
| E | 22% | 0% | 25% | 17% |
| W | 17% | 17% | 15% | 12% |
| Total | 100% | 100% | 100% | 100% |

The costs of producing a widget are $48 and $72 for the Standard and the Deluxe widget respectively. The monthly fixed costs for this year are $20000. Delivery costs depend solely on the region and are shown in this table:

| Region | North | South-East | South-West | East | West |
|---|---|---|---|---|---|
| *Unit Delivery Cost* | *$10.25* | *$9.73* | *$9.58* | *$8.26* | *$11.02* |

The company CEO wants to see the following results:

- The monthly sales amount and units per product.
- The monthly units, sales amount, costs and profit.
- The total profit.

# Acme TechnoWidget Company Formula Diagram

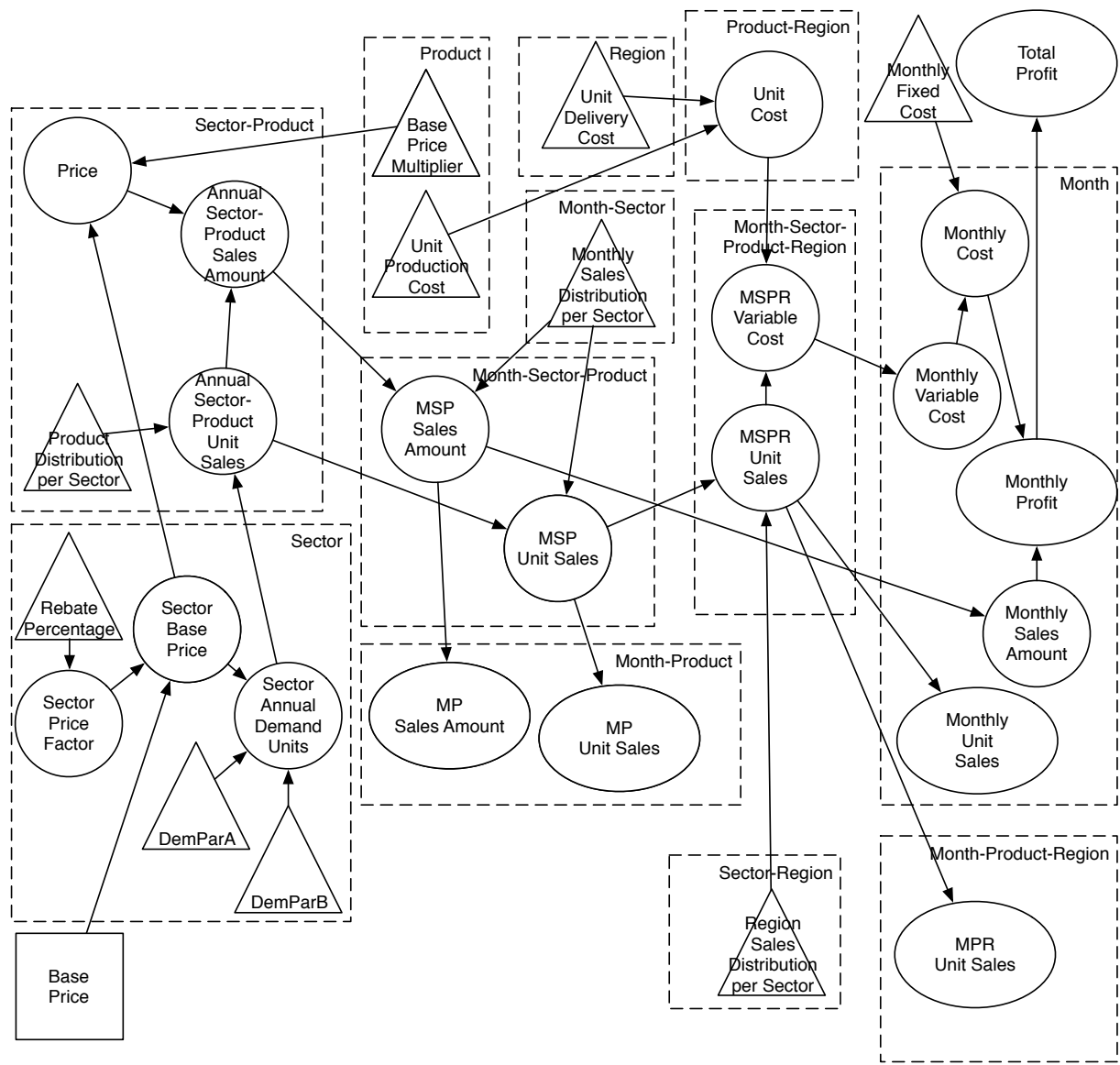

# Acme TechnoWidget Company Formula List

| No | Variable | Type | Dimension Set | Value / Formula |
|---|---|---|---|---|
| 1 | Base Price | Input | | 100 |
| 2 | Base Price Multiplier | Data | Product | (1, 1.45) |
| 3 | Unit Production Cost | Data | Product | list of values |
| 4 | Rebate Percentage | Data | Sector | list of values |
| 5 | Sector Price Factor | Calculated | Sector | 1-Rebate Percentage |
| 6 | Sector Base Price | Calculated | Sector | Base Price * Sector Price Factor |
| 7 | DemParA | Data | Sector | list of values |
| 8 | DemParB | Data | Sector | list of values |
| 9 | Sector Annual Demand Units | Calculated | Sector | DemParA*DemParB^-Sector Base Price |
| 10 | Unit Delivery Cost | Data | Region | list of values |
| 11 | PR Unit Cost | Calculated | Product-Region | Unit Production Cost + Unit Delivery Cost |
| 12 | Product Distribution per Sector | Data | Sector-Product | list of values |
| 13 | Annual Sector-Product Unit Sales | Calculated | Sector-Product | Sector Annual Demand Units * Product Distribution per Sector |
| 14 | Price | Calculated | Sector-Product | Sector Base Price * Base Price Multiplier |
| 15 | Annual Sector-Product Sales Amount | Calculated | Sector-Product | Annual Sector-Product Unit Sales * Price |

| No | Variable | Type | Dimension Set | Value / Formula |
|---|---|---|---|---|
| 16 | Region Sales Distribution per Sector | Data | Sector-Region | list of values |
| 17 | Monthly Sales Distribution per Sector | Data | Month-Sector | list of values |
| 18 | MSP Unit Sales | Calculated | Month-Sector-Product | Annual Sector-Product Unit Sales * Monthly Sales Distribution per Sector |
| 19 | MSP Sales Amount | Calculated | Month-Sector-Product | Annual Sector-Product Sales Amount * Monthly Sales Distribution per Sector |
| 20 | MSPR Unit Sales | Calculated | Month-Sector-Product-Region | MSP Unit Sales * Region Sales Distribution per Sector |
| 21 | MSPR Variable Cost | Calculated | Month-Sector-Product-Region | MSPR Unit Sales * PR Unit Cost |
| 22 | Monthly Variable Cost | Calculated | Month | SUM(MSPR Variable Cost) |
| 23 | Monthly Unit Sales | Output | Month | SUM(MSPR Unit Sales) |
| 24 | Monthly Sales Amount | Calculated | Month | SUM(MSP Sales Amount) |
| 25 | Monthly Fixed Cost | Data | | |
| 26 | Monthly Costs | Calculated | Month | Monthly Fixed Cost + Monthly Variable Cost |
| 27 | Monthly Profit | Calculated | Month | Monthly Sales Amount - Monthly Costs |
| 28 | MPR Unit Sales | Output | Month-Product-Region | SUM(MSPR Unit Sales) |
| 29 | MP Unit Sales | Output | Month-Product | SUM(MSP Unit Sales) |
| 30 | MP Sales Amount | Output | Month-Product | SUM(MSP Sales Amount) |
| 31 | Total Profit | Output | | SUM(Monthly Profit) |